# Automatic Cloud Resource Scaling Algorithm based on Long Short-Term Memory Recurrent Neural Network

Ashraf A. Shahin[1,2]
[1]College of Computer and Information Sciences,
Al Imam Mohammad Ibn Saud Islamic University (IMSIU)
Riyadh, Kingdom of Saudi Arabia
[2]Department of Computer and Information Sciences, Institute of Statistical Studies & Research,
Cairo University,
Cairo, Egypt

***Abstract*—Scalability is an important characteristic of cloud computing. With scalability, cost is minimized by provisioning and releasing resources according to demand. Most of current Infrastructure as a Service (IaaS) providers deliver threshold-based auto-scaling techniques. However, setting up thresholds with right values that minimize cost and achieve Service Level Agreement is not an easy task, especially with variant and sudden workload changes. This paper has proposed dynamic threshold based auto-scaling algorithms that predict required resources using Long Short-Term Memory Recurrent Neural Network and auto-scale virtual resources based on predicted values. The proposed algorithms have been evaluated and compared with some of existing algorithms. Experimental results show that the proposed algorithms outperform other algorithms.**

## I. Introduction

One of the important features provided by cloud computing is Scalability, which is the ability to scale allocated computational resources on-demand [1]. Scalability feature allows users to run their applications in an elastic manner, use only computational resources they need, and pay only for what they use. However, the process of instantiating new virtual machines takes 5-15 minutes [2]. Therefore, predicting future demand might be required to deal with variable demands and being able to scale in advance. In the current literature, many diverse auto-scaling techniques have been proposed to scale computational resources according to predicted workload [3, 4, 5, 6].

However, one of the most famous problems that face current auto-scaling techniques is Slashdot problem; where auto-scaling technique might not be able to scale in case of sudden influx of valid traffic. Slashdot is unpredictable flash-crowd workload. Flash-crowd workload reduces cloud service providers' revenue by violating Service Level Agreement.

Slashdot effects can be reduced by detecting Slashdot situations at earlier stages and performing appropriate scaling actions. However, detecting Slashdot situations at earlier stages is not an easy task. Even if Slashdot is detected, finding suitable scaling action is a very hard task. Recently, several machine-learning techniques (e.g. Support Vector Machine, Neural Networks, and Linear Regression) have been used to predict cloud workload [7, 8, 9]. However, most of currently used techniques cannot remember events if there are very long and variant time lags between events, as in Slashdot.

To improve memorization of standard feed forward neural network, Jeff Elman has proposed recurrent neural network (RNN), which extends standard feed forward neural network by adding internal memory [10]. RNNs can learn when the gap between relevant events is small (less than 10-step time lags). Unfortunately, conventional RNNs still unable to learn when gap between relevant events grows [1]. In 1997, Hochreiter & Schmidhuber have proposed a special type of RNN, called Long Short-Term Memory network (LSTM), with ability to recognize and learn long-term dependencies (up to 1000-step time lags between relevant events)[1].

This paper tries to answer the question: can we reduce Slashdot effects by using LSTM-RNN? To answer this question, this paper has proposed two auto-scaling algorithms. The first algorithm avoids long and variant time lags between Slashdot situations by using two different LSTM-RNNs. The first LSTM-RNN is employed to deal with normal workload while the second LSTM-RNN is exploited to deal with Slashdot workload. The second algorithm investigates applicability of using one LSTM-RNN to deal with both normal and Slashdot workloads. Performance of the proposed algorithms have been evaluated and compared with some of existing algorithms using CloudSim with real traces. Experimental results show that the first auto-scaling algorithm, which uses two LSTM-RNNs, outperforms other algorithms.

The rest of this paper is structured as follows. Section 2 gives a brief background on Long Short-Term Memory recurrent neural network (LSTM-RNN). Section 3 overviews related work in the area of automatic cloud resources scaling. Section 4 briefly describes the proposed algorithms. Following this, Section 5 evaluates performance of the proposed algorithms using CloudSim simulator with real workloads and compares their performance with some of existing algorithms. Finally, Section 6 concludes.

## II. LSTM-RNN

Feed forward neural network is a set of connected neurons that try to capture and represent underlying relationships in a set of data [10]. One of the major limitations of feed forward neural network is that it does not consider order in time and only remember few moments of training from their recent past. Therefore, feed forward neural network cannot recognize or learn sequential or time-varying patterns [10].

Alternatively, recurrent neural networks (RNN) determine new response by using feedback loops, which combine current inputs with outputs of the previous moment. Feedback loops allow sequential information to persist and allow recurrent networks to perform tasks that cannot be performed by feed forward neural networks [1].

Figure 1 shows simple recurrent neural network design, which was proposed by Elman. New layer (called *context layer*) has been added to standard feed forward neural network. Context units receive inputs from, and return their results to hidden units. Context units allow RNN to memorize its previous state [10].

Unfortunately, regular RNN still loses its memory very fast. In 1997, Hochreiter & Schmidhuber have proposed a special type of RNN, called Long Short-Term Memory network (LSTM), with ability to recognize and learn long-term dependencies. Long Short-Term Memory blocks have been added to the hidden layers of RNN [11]. As shown in Fig. 2, each memory block contains memory cell to store internal state and contains three different types of gates (input, output and forget gates) to manage cell state and output using activation function (usually sigmoid). The input gate decides what information to store in the memory cell. The output gate decides when to read information from the memory cell. The forget gate decides how long to store information in the memory cell. In 2002, Schmidhuber et al. have enhanced memory block by adding peephole connections from its internal cell to its gates. Peephole connections allow LSTM to learn precise timing between relevant events [1].

## III. Related Work

Recently, several auto-scaling techniques have been proposed. In [12], Gandhi et al. have proposed auto-scaling approach, called Dependable Compute Cloud, to scale infrastructure automatically without accessing application-level and without offline application profiling. The proposed approach proactively scales application deployment based on monitoring information from resource-level and based on performance requirements that are specified by users. Multi-tier cloud application is approximated using product-form queueing-network model. Kalman filtering technique is employed to predict required parameters without accessing user's application. However, the proposed approach has not considered Slashdot and has assumed that incoming requests have Poisson arrivals.

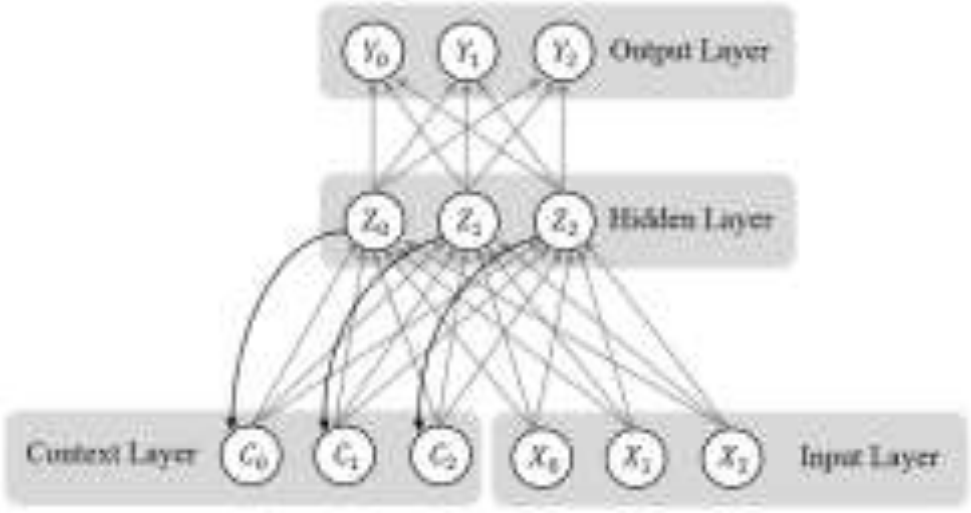

Fig. 1. Simple Recurrent Neural Network Design

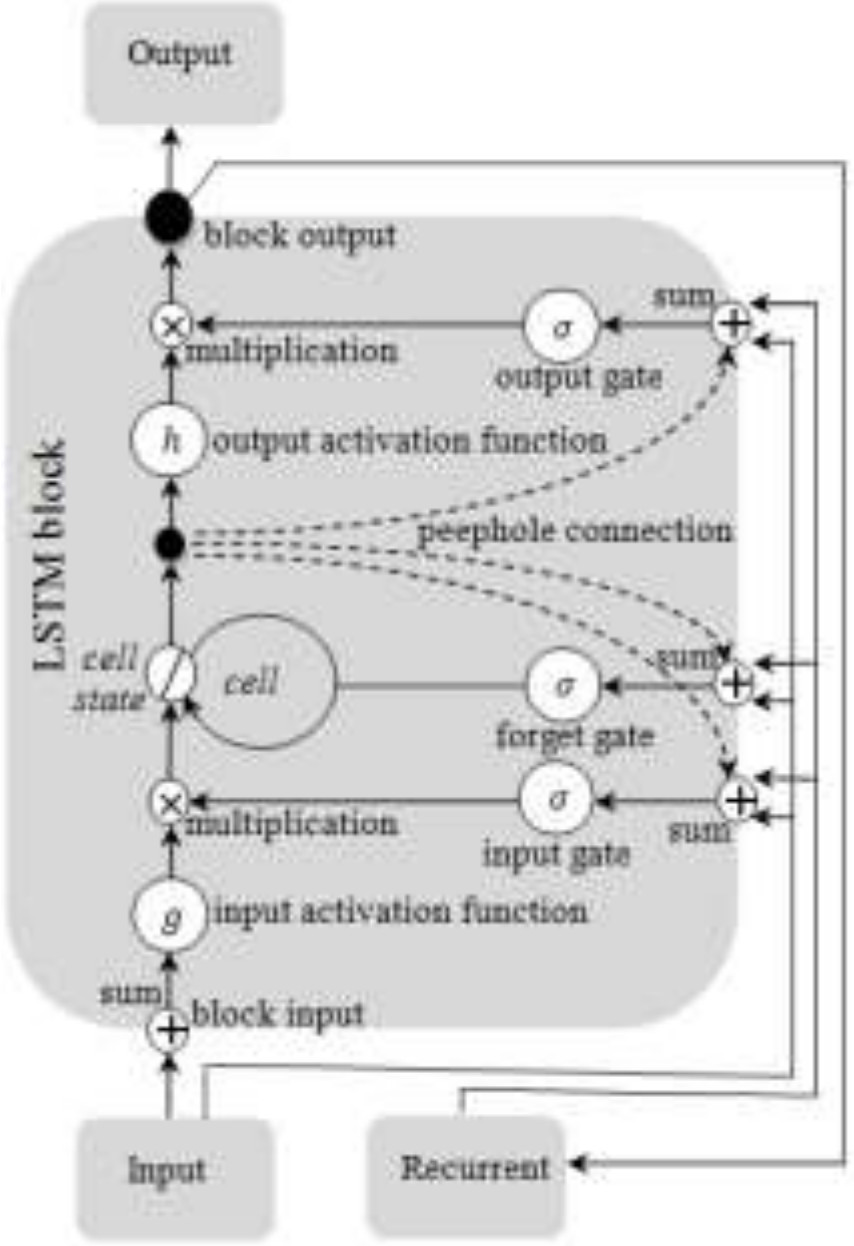

Fig. 2. Long Short-Term Memory block [11]

In [13], Moore et al. proposed a hybrid elasticity controller that coordinates between reactive and predictive scalability controllers to enhance cloud applications scalability. Both controllers act concurrently. Cloud applications' administrators configure scaling rules, which are monitored by reactive controller. After some condition has already been met, reactive controller submits scaling requests to centralized decision manager. If the predictive controller is certain of what action to take then it submits scaling requests to centralized decision manager. Otherwise, the predictive controller continues to learn. Decision manager receives, validates, and executes all triggered scaling requests. Although, performance of the proposed elasticity controller has been evaluated using two real traces (ClarkNet web server trace logs and FIFA 1998 World Cup Access logs), none of these traces has Slashdot. Therefore, performance of the proposed elasticity controller has not been evaluated with Slashdot.

Lin et al. [3, 6] proposed auto-scaling system, which monitors incoming requests and HTTP response time to recognize cloud applications' performance. Auto-scaling algorithm was proposed based on recognized performance.

Furthermore, Lin et al. proposed an algorithm to analyze the workload trend to reduce the number of peaks in response time caused by the variability of workload. Although, the authors have mentioned that the proposed scaling strategy can respond to variant and sudden workload changes in short time, the proposed strategy has not been evaluated using sudden workload changes and only evaluated using short workload (200 minutes) with predictable seasonality.

Kanagala and Sekaran [14] have proposed Threshold-based auto-scaling approach, which minimizes violation of service level agreement by considering virtual machine turnaround time and virtual machine stabilization time during adapting thresholds. Thresholds are dynamically specified by using double exponential smoothing. To set upper threshold, double exponential smoothing is used to predict at which time the system will reach max load and specify point before this time to be used as upper threshold. To scale down, double exponential smoothing is used to predict point before reaching the minimum system load and use it as lower threshold. However, weights that are assigned to observations by double exponential smoothing method are decreased exponentially while observations get older. Therefore, double exponential smoothing method does not able to remember Slashdot when there are long time lags.

Mao et al. [4, 5, 15] proposed auto-scaling mechanism, which considers both user performance requirements and cost concerns. Performance requirements are specified by assigning soft deadline for each job. The proposed auto-scaling mechanism allocates/deallocates virtual machines and schedules tasks on virtual machines to finish each job within its deadline with minimum cost. However, instantiating new VMs requires at least 10 minutes. Thus, probability of violating Service Level Agreement is increased.

Nikravesh et al. [16] proposed a proactive auto-scaling system based on Hidden Markov Model. Their experiments shown that scaling decisions that are generated using Hidden Markov Model are more accurate than scaling decisions that are generated using support vector machine, neural networks, and linear regression. In [7, 8, 9], Bankole and Ajila have applied three machine-learning techniques: Support Vector Machine, Neural Networks, and Linear Regression to proactively scale provisioned cloud resources for multitier web applications. Their results show that Support Vector Machine outperforms other techniques in predicting future resource demands. Although, several auto-scaling techniques have been proposed during the last few years, most of them do not consider Slashdot.

## IV. Proposed Algorithms

As shown in Algorithm 1, inputs are as following. The first input, $CPU_H$, is the history of total required CPU. Total required CPU at time $t$, $CPU_H(t)$, is calculated as sum of all required CPU for coming requests at time $t$.

To enhance prediction accuracy of the proposed algorithms, sliding window technique is utilized. Sliding window has been used in many areas to improve prediction accuracy [2]. The input $\mathrm{W_{length}}$ specifies size of sliding window that will be used during prediction.

The input $\mathrm{VM_{delay}}$ represents delay of starting up new VM. $CPU_{PH1}$ and $CPU_{PH2}$ are history of previously predicted CPU by using first and second LSTM-RNN respectively. $\mathrm{MAPE}_{Normal}$ and $MAPE_{Slashdot}$ represent prediction accuracy of first and second LSTM-RNN respectively. Prediction accuracy is calculated as Mean Absolute Percentage Error (MAPE).

The first auto-scaling algorithm uses two different LSTM-RNNs for forecasting future demand. The first LSTM-RNN is trained by normal workload without Slashdot and the second LSTM-RNN is trained with Slashdot workload only. $\mathrm{MAPE}_{Normal}$ and $MAPE_{Slashdot}$ are continuously updated using predicted and observed CPU. Required CPU after $\mathrm{VM_{delay}}$ step-ahead is forecasted by using LSTM-RNN with lowest MAPE. Predicted CPU is sent to Scaling Decision Maker algorithm to decide appropriate scaling action. Number of VMs to scale up or down is specified according to the difference between predicted and provisioned resources after $\mathrm{VM_{delay}}$ step-ahead.

Algorithm 2 shows steps of the second auto-scaling algorithm, which uses only one LSTM-RNN to predict required CPU with normal and Slashdot workloads.

ALGORITHM 1: Auto-scaling with two LSTM-RNN

**INPUTS**:
$CPU_H$: history of total required CPU
$\mathrm{W_{length}}$: sliding window length
$\mathrm{VM_{delay}}$: VM startup delay
$CPU_{PH1}$: history of predicted CPU using $LSTM_RNN_{Normal}$
$CPU_{PH2}$: history of predicted CPU using $LSTM_RNN_{Slashdot}$
$\mathrm{MAPE}_{Normal}$: prediction accuracy of $LSTM_RNN_{Normal}$
$\mathrm{MAPE}_{Slashdot}$: prediction accuracy of $LSTM_RNN_{Slashdot}$
**OUTPUTS**:
Scaling decision
**Begin**
1: $w_t =$ Get sliding window from $CPU_H$ with length $\mathrm{W_{length}}$
2: $CPU_{PH1}(t) =$ Predict required CPU after $\mathrm{VM_{delay}}$ step-ahead using $LSTM_RNN_{Normal}(w_t)$
3: $CPU_{PH2}(t) =$ Predict required CPU after $\mathrm{VM_{delay}}$ step-ahead using $LSTM_RNN_{Slashdot}(w_t)$
4: Update $\mathrm{MAPE}_{Normal}$ and $\mathrm{MAPE}_{Slashdot}$
5: **if** $\mathrm{MAPE}_{Normal} < \mathrm{MAPE}_{Slashdot}$
6: Call Scaling Decision Maker using $CPU_{PH1}(t)$
7: **else**
8: Call Scaling Decision Maker using $CPU_{PH2}(t)$
9: **endif**
10: **return** scaling decision
**End**

ALGORITHM 2: Auto-scaling with one LSTM-RNN

**INPUTS**:
$CPU_H$: history of total required CPU
$W_{length}$: sliding window length
$VM_{delay}$: VM startup delay
$CPU_{PH}$: history of predicted CPU
**OUTPUTS**:
Scaling decision
**Begin**
$w_t =$ Get sliding window from $CPU_H$ with length $W_{length}$
$CPU_{PH}(t) =$ Predict required CPU after $VM_{delay}$ step-ahead
Call Scaling Decision Maker using $CPU_{PH}(t)$
**return** scaling decision
**End**

Scaling decision maker algorithm is shown in Algorithm 3. Scaling decision maker algorithm uses three thresholds: upper threshold *ThrU*, lower threshold *ThrL*, and *ThrbU*, which is slightly below the upper threshold *ThrU*. If required CPU crosses above *ThrU*, virtual resources are considered over utilized and have to be scaled up. If required CPU crosses above *ThrbU* and does not cross above *ThrU* for a pre-specified number of times, virtual resources are considered over utilized and virtual resources have to be scaled up. In another hand, if required CPU crosses below *ThrL* for a pre-specified number of times, virtual resources are considered underutilized and some virtual resources have to be released.

Thresholds are initialized by the same values for all applications. However, due to variation nature of workloads, setting the same values for all applications increases the probability of violating service level agreements. Therefore, all thresholds are periodically and automatically adapted using Median Absolute Deviation of required CPU history for each application.

$$ThrU = 1 - c_1 . MAD,$$

$$ThrbU = 1 - c_2 . MAD,$$

$$ThrL = 1 - c_3 . MAD,$$

wher $c_1, c_2, and\ c_3 \in \mathbb{R}^+, such\ that\ c_1 < c_2 < c_3$, and $MAD$ is median of absolute deviations from median of required CPU. Using $c_1, c_2, and\ c_3$, we can adapt the safety of the proposed algorithm. For example, lower values for $c_1, and\ c_2$ decrease the cost, but increase the probability of violating service level agreements.

## V. Performance Evaluation

Proposed algorithms have been implemented using Cloudsim simulator with deep-learning library called Deeplearning4j [20]. Performances of the proposed algorithms have been compared with two auto-scaling approaches, which are proposed by Kanagala et al. [14], and Hasan et al. [17]. The following subsections describe evaluation environment settings and discuss simulations' results.

### A. Evaluation environment settings

The proposed algorithms have been evaluated using CloudSim simulator with real trace called NASA Log [18]. NASA Log contains two month's *HTTP* requests to the *NASA Kennedy Space Center WWW server*, which is located in Florida. This log was collected from 00:00:00 July 1, 1995 to 23:59:59 July 31, 1995 and from 00:00:00 August 1, 1995 to 23:59:59 August 31, 1995. Fig. 3 shows number of requests that are generated according to NASA Log from August 1 to August 31.

Slashdot has been added to NASA Log from [19], which contains number of hits for July 26 2000; the day the AUUG/LinuxSA InstallFest story hit Slashdot (Fig. 4 shows number of requests versus time). Fig. 5 shows NASA Log after adding Slashdot.

ALGORITHM 3: Scaling Decision Maker

**INPUTS**:
$CPU_H$: history of total required CPU
$CPU_P(t)$: predicted CPU at time $t$
$Scaling_{delay}$: duration before scaling
**OUTPUTS**:
Scaling Action
**Begin**
$MAD =$ Get Median Absolute Deviation of $CPU_H$
$ThrU = 1 - c_1 * MAD$
$ThrbU = 1 - c_2 * MAD$
$ThrL = 1 - c_3 * MAD$
$Tick_Up_Timer = 0$
$Tick_Down_Timer = 0$
**if** $(CPU_P(t) > ThrU)$
  $Scaling_Action = scale_Up$
  $Tick_Down_Timer = 0$
  $Tick_Up_Timer = 0$
**else**
  **if**$(CPU_P(t) > ThrbU)$
    $Tick_Down_Timer = 0$
    $Tick_Up_Timer{+}{+}$
    **if**$(Tick_Up_Timer > Scaling_{delay})$
      $Scaling_Action = scale_Up$
    **endif**
  **else**
    **if** $(CPU_P(t) < ThrL)$
      $Tick_Up_Timer = 0$
      $Tick_Down_Timer{+}{+}$
      **if**$(Tick_Down_Timer > Scaling_{delay})$
        $Scaling_Action = scale_Down$
      **endif**
    **else**
      $Tick_Down_Timer = 0$
      $Tick_Up_Timer = 0$
    **endif**
  **endif**
**endif**
**return** Scaling_Action
**End**

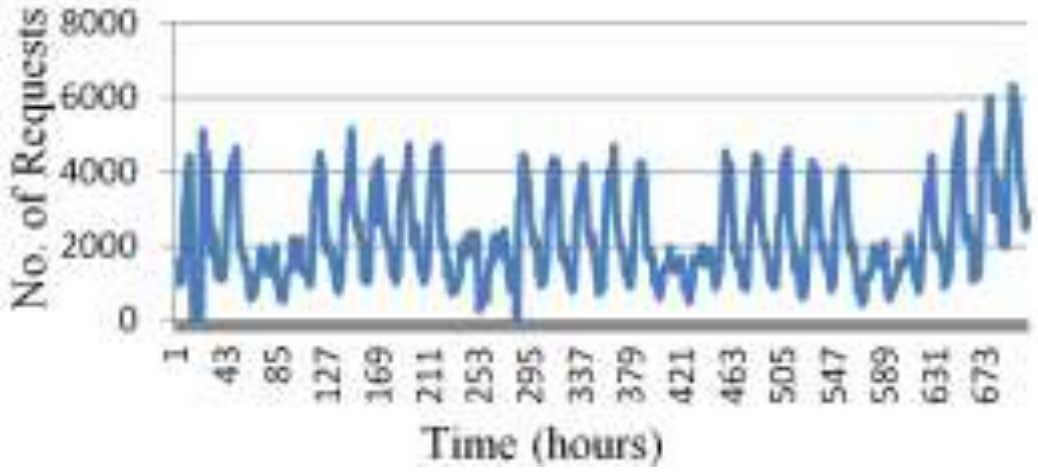

Fig. 3. Generated requests according NASA Log from August 1 to August 31

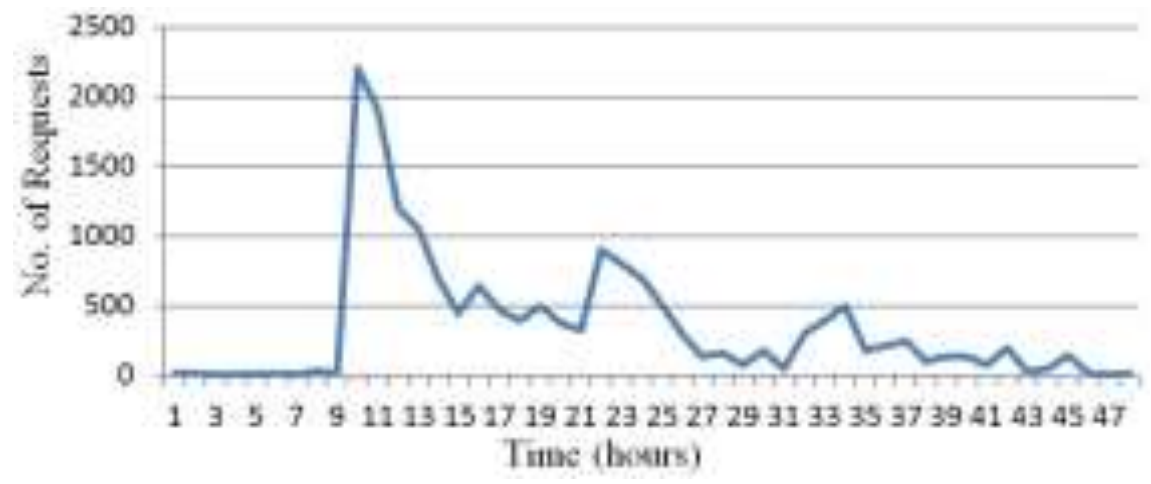

Fig. 4. Number of requests versus time for Slashdot

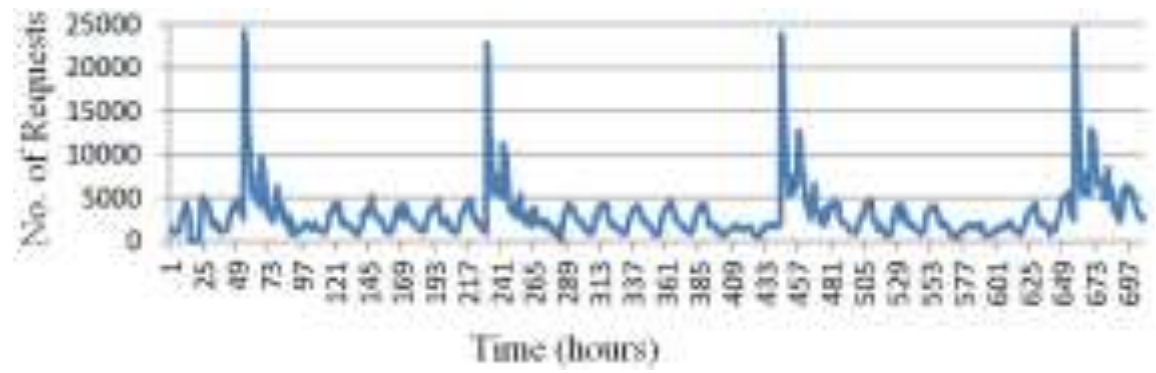

Fig. 5. NASA Log with Slashdot

To implement LSTM-RNN, *Deeplearning4j* library has been used. *Deeplearning4j* is an open-source deep-learning library in Java. *Deeplearning4j* is developed by San Francisco-based business intelligence and enterprise software firm [20].

## B. *Evaluation results*

Fig. 6 and Table 1 show number of running VMs during period from hour 221 to hour 248, which contains the second Slashdot (as shown in Fig. 5). The proposed algorithms increase number of running VMs (between 221 and 230) among other approaches to deal with Slashdot and rapidly decrease number of running VMs (between 230 and 250) to minimize cost.

Fig. 6 and Table 1 show that number of provisioned VMs by the proposed algorithms is higher than provisioned VMs by the related approaches. These VMs are incorporated to achieve large number of requests in short response time as shown in Fig. 7, Table 2, Fig. 8, and Table 3.

In [17], fixed number of VMs is defined to be allocated or de-allocated during scaling up or down. This fixed number limits scaling speed through Slashdot. In the proposed algorithms, number of VMs is variant and depends on growth or decrease of the workload.

In [14] and [17], if workload goes across the upper threshold for a pre-specified duration, they start to scale up. During this period, Service Level Agreement (SLA) will be violated and some penalty has to be incurred by providers. Moreover, duration of SLA violation will be extended to include startup delay of new VMs, which sometimes takes around 10 minutes. In the proposed algorithms, VMs will be scaled up directly if predicted workload goes across the upper threshold. Therefore, the proposed algorithms act faster to provide enough resources to achieve coming requests.

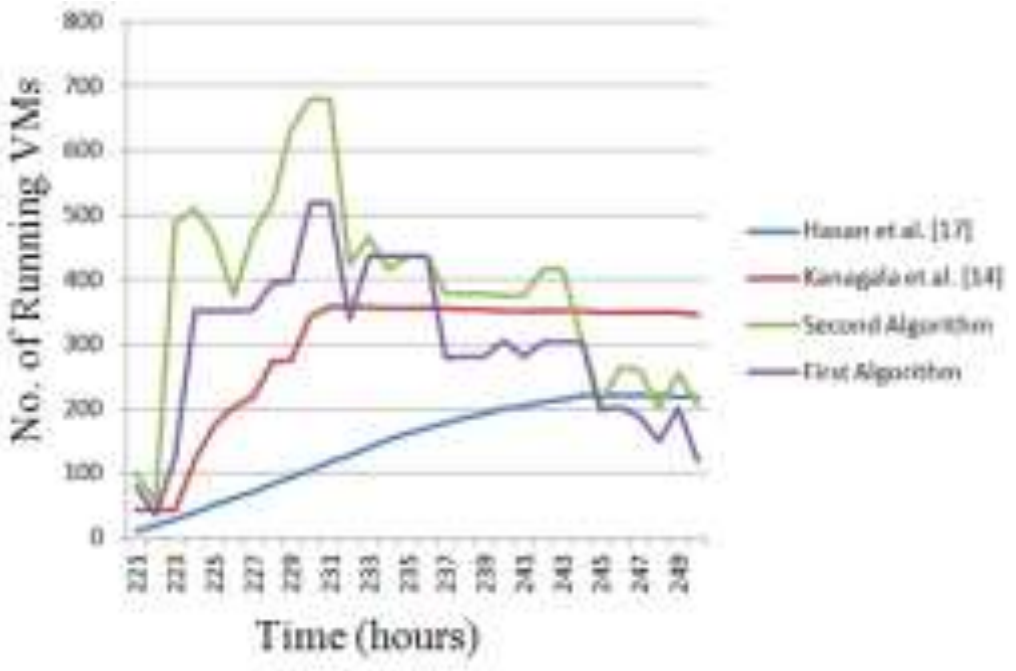

Fig. 6. Number of running VMs versus time

TABLE I. NUMBER OF RUNNING VMS VERSUS TIME

| Time (hour) | First Algorithm | Second Algorithm | Kanagala et al. [14] | Hasan et al. [17] |
|---|---|---|---|---|
| 221 | 79 | 100 | 43 | 11 |
| 222 | 36 | 55 | 43 | 19 |
| 223 | 125 | 488 | 44 | 28 |
| 224 | 352 | 510 | 119 | 40 |
| 225 | 352 | 466 | 173 | 51 |
| 226 | 352 | 376 | 203 | 62 |
| 227 | 353 | 472 | 219 | 72 |
| 228 | 396 | 520 | 272 | 83 |
| 229 | 397 | 633 | 274 | 94 |
| 230 | 519 | 680 | 344 | 106 |
| 231 | 519 | 680 | 357 | 117 |
| 232 | 339 | 424 | 357 | 128 |
| 233 | 435 | 466 | 357 | 139 |
| 234 | 435 | 416 | 356 | 151 |
| 235 | 435 | 435 | 356 | 161 |
| 236 | 435 | 435 | 356 | 169 |
| 237 | 279 | 378 | 355 | 178 |
| 238 | 280 | 378 | 353 | 187 |
| 239 | 281 | 378 | 353 | 193 |
| 240 | 305 | 376 | 352 | 201 |
| 241 | 281 | 376 | 352 | 205 |
| 242 | 303 | 416 | 352 | 211 |
| 243 | 303 | 415 | 352 | 215 |
| 244 | 303 | 305 | 351 | 220 |
| 245 | 199 | 209 | 350 | 220 |
| 246 | 202 | 262 | 349 | 220 |
| 247 | 188 | 261 | 349 | 220 |
| 248 | 150 | 200 | 349 | 220 |
| 249 | 200 | 254 | 349 | 219 |
| 250 | 120 | 207 | 346 | 218 |

In [17], they scale down if the trend is down even if the load does not cross the lower threshold, which means that VMs will be shrunken even if we do not need that. Moreover, in [17], it terminates VM after marking it to be terminated after 5 minutes even if it is already finished, which sometimes increases the cost if these few minutes add more hour cost. In addition, it can increase SLA violation if there are running requests need more time.

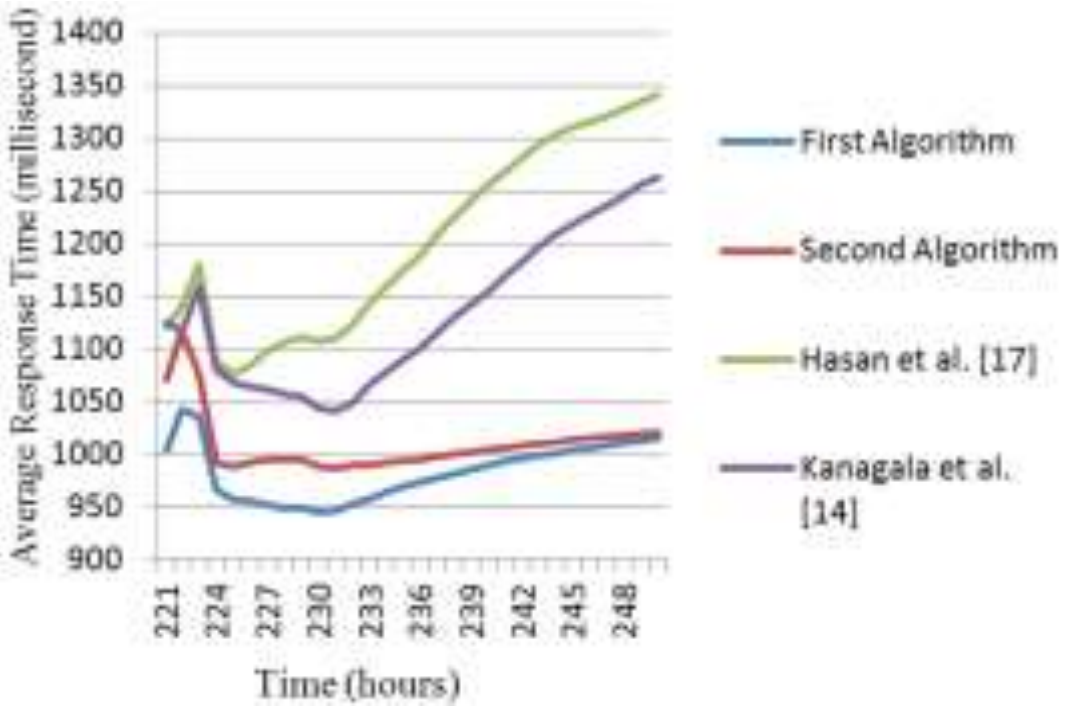

Fig. 7. Average response time

TABLE II. AVERAGE RESPONSE TIME (MILLISECOND)

| Time (hour) | First Algorithm | Second Algorithm | Kanagala et al. [14] | Hasan et al. [17] |
|---|---|---|---|---|
| 221 | 1003 | 1071 | 1125 | 1121 |
| 222 | 1042 | 1117 | 1116 | 1139 |
| 223 | 1036 | 1076 | 1159 | 1181 |
| 224 | 967 | 993 | 1080 | 1089 |
| 225 | 958 | 988 | 1068 | 1077 |
| 226 | 956 | 993 | 1064 | 1084 |
| 227 | 952 | 995 | 1061 | 1098 |
| 228 | 949 | 996 | 1057 | 1106 |
| 229 | 950 | 996 | 1055 | 1111 |
| 230 | 946 | 988 | 1044 | 1108 |
| 231 | 947 | 988 | 1042 | 1110 |
| 232 | 953 | 990 | 1050 | 1123 |
| 233 | 958 | 990 | 1065 | 1143 |
| 234 | 964 | 992 | 1078 | 1159 |
| 235 | 969 | 993 | 1090 | 1175 |
| 236 | 974 | 996 | 1100 | 1189 |
| 237 | 978 | 998 | 1116 | 1209 |
| 238 | 982 | 1000 | 1130 | 1226 |
| 239 | 986 | 1003 | 1143 | 1241 |
| 240 | 990 | 1005 | 1155 | 1256 |
| 241 | 994 | 1007 | 1169 | 1269 |
| 242 | 996 | 1008 | 1184 | 1282 |
| 243 | 999 | 1010 | 1197 | 1293 |
| 244 | 1001 | 1012 | 1208 | 1304 |
| 245 | 1004 | 1014 | 1218 | 1310 |
| 246 | 1006 | 1016 | 1228 | 1316 |
| 247 | 1009 | 1017 | 1236 | 1321 |
| 248 | 1011 | 1018 | 1245 | 1327 |
| 249 | 1013 | 1019 | 1256 | 1335 |
| 250 | 1015 | 1021 | 1263 | 1341 |

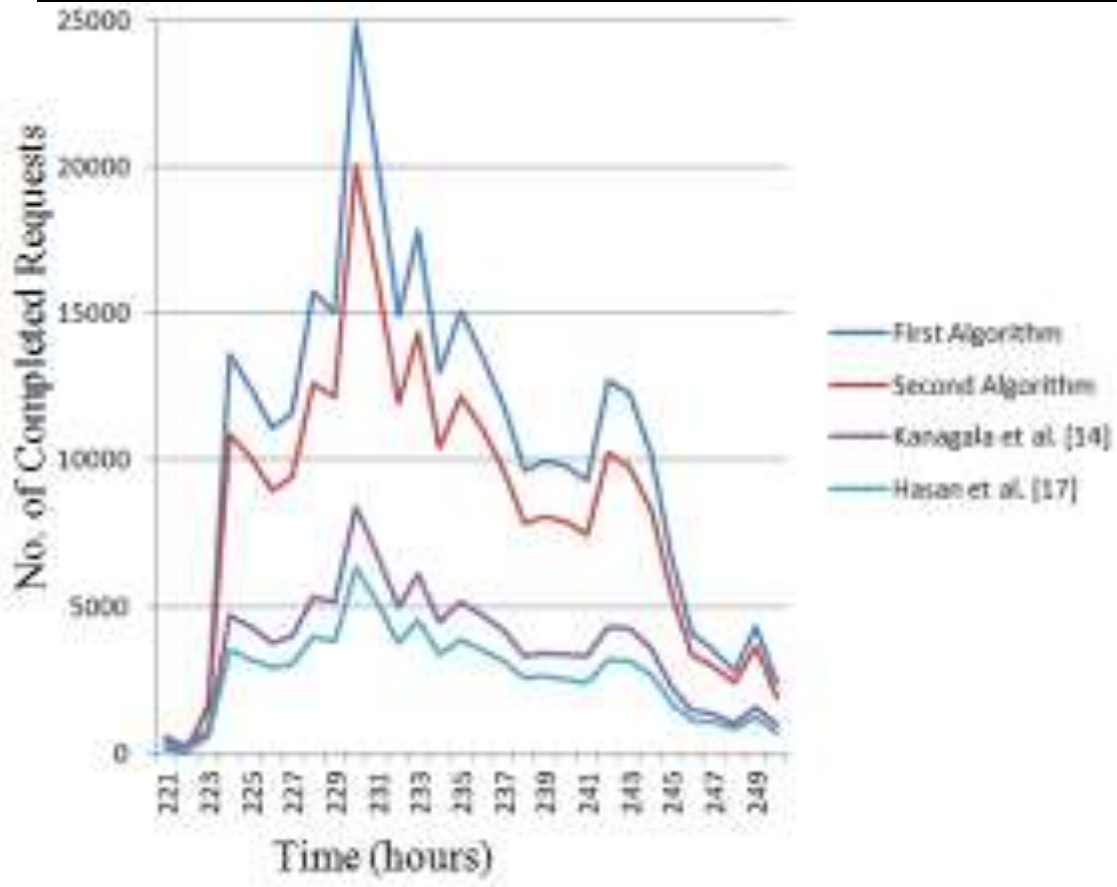

Fig. 8. Number of completed requests

## VI. CONCLUSION

Although, elasticity is one of cloud computing cornerstones that attract many companies to host their applications in the cloud, most of current dynamic resource scaling systems do not have ability to deal with Slashdot. Slashdot prevents companies from gaining benefits of cloud computing elasticity and increases the probability of losing customers. Motivated by this problem, this paper has proposed two auto-scaling algorithms based on Long Short-Term Memory Recurrent Neural Network to minimize Slashdot effects.

TABLE III. NUMBER OF COMPLETED REQUESTS

| Time (hour) | First Algorithm | Second Algorithm | Kanagala et al. [14] | Hasan et al. [17] |
|---|---|---|---|---|
| 221 | 30 | 530 | 412 | 323 |
| 222 | 155 | 92 | 91 | 153 |
| 223 | 1908 | 1646 | 649 | 585 |
| 224 | 13456 | 10992 | 4658 | 3513 |
| 225 | 12597 | 10142 | 4305 | 3266 |
| 226 | 11049 | 8884 | 3776 | 2881 |
| 227 | 11543 | 9471 | 3995 | 3008 |
| 228 | 15729 | 12508 | 5393 | 4039 |
| 229 | 15151 | 12154 | 5159 | 3891 |
| 230 | 25011 | 20060 | 8517 | 6335 |
| 231 | 20104 | 16158 | 6759 | 5145 |
| 232 | 14872 | 12034 | 5060 | 3754 |
| 233 | 17741 | 14364 | 6115 | 4503 |
| 234 | 12797 | 10316 | 4369 | 3377 |
| 235 | 15011 | 12175 | 5119 | 3838 |
| 236 | 13742 | 11056 | 4759 | 3521 |
| 237 | 11873 | 9612 | 4159 | 3108 |
| 238 | 9758 | 7937 | 3290 | 2500 |
| 239 | 9832 | 8067 | 3406 | 2570 |
| 240 | 9833 | 7887 | 3386 | 2566 |
| 241 | 9187 | 7582 | 3147 | 2355 |
| 242 | 12763 | 10144 | 4353 | 3281 |
| 243 | 12247 | 9877 | 4097 | 3174 |
| 244 | 10265 | 8252 | 3607 | 2649 |
| 245 | 6397 | 5350 | 2278 | 1742 |
| 246 | 4072 | 3374 | 1452 | 1115 |
| 247 | 3535 | 2948 | 1347 | 1047 |
| 248 | 2903 | 2356 | 1040 | 766 |
| 249 | 4429 | 3634 | 1591 | 1224 |
| 250 | 2340 | 2052 | 874 | 735 |

The proposed algorithms have been empirically evaluated against some of existing approaches. Experiment results have showed that the proposed algorithms outperform others on both cost and service level agreement. Based on these results, this paper concludes that using Long Short-Term Memory Recurrent Neural Network to recognize and deal with Slashdot can minimizes it effects.

In the future, deep Long Short-Term Memory Recurrent Neural Network will be exploited to recognize Slashdot behavior. Deep LSTM-RNN has been effectively applied in many areas and has proved its efficiency throughout the years. Deep LSTM-RNN offers more benefits over standard LSTM RNNs by having several hidden layers. Each layer processes some part of the task before sending it to the next layer.

## REFERENCES

[1] J. S. Felix A. Gers, Nicol N. Schraudolph, “Learning precise timing with lstm recurrent networks,” *Journal of Machine Learning Research*,

vol. 3, pp. 115–143, 2002. [Online]. Available: http://www.jmlr.org/-papers/volume3/gers02a/gers02a.pdf

[2] S. Islam, J. Keung, K. Lee, and A. Liu, "Empirical prediction models for adaptive resource provisioning in the cloud," *Future Gener. Comput. Syst.*, vol. 28, no. 1, pp. 155–162, Jan. 2012. [Online]. Available: http://-dx.doi.org/10.1016/j.future.2011.05.027

[3] C.-C. Lin, J.-J. Wu, P. Liu, J.-A. Lin, and L.-C. Song, "EnglishAutomatic resource scaling for web applications in the cloud," in *EnglishGrid and Pervasive Computing*, ser. Lecture Notes in Computer Science, J. Park, H. Arabnia, C. Kim, W. Shi, and J.-M. Gil, Eds. Springer Berlin Heidelberg, vol. 7861, pp. 81–90, 2013. [Online]. Available: http://dx.doi.org/10.1007/978-3-642-38027-3_9

[4] M. Mao and M. Humphrey, "Auto-scaling to minimize cost and meet application deadlines in cloud workflows," in *2011 International Conference for High Performance Computing, Networking, Storage and Analysis (SC)*, pp. 1–12, Nov 2011.

[5] M. Mao, J. Li, and M. Humphrey, "Cloud auto-scaling with deadline and budget constraints," in *2010 11th IEEE/ACM International Conference on Grid Computing (GRID)*, pp. 41–48, Oct 2010.

[6] C.-C. Lin, J.-J. Wu, J.-A. Lin, L.-C. Song, and P. Liu, "Automatic resource scaling based on application service requirements," in *2012 IEEE 5th International Conference on Cloud Computing (CLOUD)*, pp. 941–942, June 2012.

[7] A. Bankole and S. Ajila, "Cloud client prediction models for cloud resource provisioning in a multitier web application environment," in *2013 IEEE 7th International Symposium on Service Oriented System Engineering (SOSE)*, pp. 156–161, March 2013.

[8] S. Ajila and A. Bankole, "Cloud client prediction models using machine learning techniques," in *2013 IEEE 37th Annual Computer Software and Applications Conference (COMPSAC)*, pp. 134–142, July 2013.

[9] A. Bankole and S. Ajila, "Predicting cloud resource provisioning using machine learning techniques," in *2013 26th Annual IEEE Canadian Conference on Electrical and Computer Engineering (CCECE)*, pp. 1–4, May 2013.

[10] J. S. Sepp Hochreiter, "Long short-term memory," *Neural Computation*, vol. 9(8), pp. 1735–1780, 1997. [Online]. Available: http://-deeplearning.cs.cmu.edu/pdfs/Hochreiter97_lstm.pdf

[11] F. B. Hasim Sak, Andrew Senior, "Long short-term memory recurrent neural network architectures for large scale acoustic modeling," in *INTERSPEECH 2014, 15th Annual Conference of the International Speech Communication Association, Singapore, September 14-18*, pp. 338–342, 2014. [Online]. Available: http://static.googleusercontent.com/media/research.google.com/en//pubs/-archive/43905.pdf

[12] A. Gandhi, P. Dube, A. Karve, A. Kochut, and L. Zhang, "Adaptive, model-driven autoscaling for cloud applications," in *11th International Conference on Autonomic Computing (ICAC 14)*. Philadelphia, PA: USENIX Association, pp. 57–64, Jun. 2014. [Online]. Available: https://www.usenix.org/conference/icac14/technical-sessions/-presentation/gandhi

[13] K. B. Laura R. Moore and T. Ellahi, "A coordinated reactive and predictive approach to cloud elasticity," in *CLOUD COMPUTING 2013 : The Fourth International Conference on Cloud Computing, GRIDs, and Virtualization*, 2013.

[14] K. Kanagala and K. Sekaran, "An approach for dynamic scaling of resources in enterprise cloud," in *2013 IEEE 5th International Conference on Cloud Computing Technology and Science (CloudCom)*, vol. 2, pp. 345–348, Dec 2013.

[15] Y. Wadia, R. Gaonkar, and J. Namjoshi, "Portable autoscaler for managing multi-cloud elasticity," in *2013 International Conference on Cloud Ubiquitous Computing Emerging Technologies (CUBE)*, pp. 48–51, Nov 2013.

[16] A. Nikravesh, S. Ajila, and C.-H. Lung, "Cloud resource auto-scaling system based on hidden markov model (hmm)," in *2014 IEEE International Conference on Semantic Computing (ICSC)*, pp. 124–127, June 2014.

[17] M. Z. Hasan, E. Magana, A. Clemm, L. Tucker, and S. L. D. Gudreddi, "Integrated and autonomic cloud resource scaling," in *2012 IEEE Network Operations and Management Symposium*, pp. 1327–1334, April 2012. DOI:10.1109/NOMS.2012.6212070

[18] Nasa-http, two months of http logs from NASA Kennedy Space Center WWW server in Florida, USA. [online] http://ita.ee.lbl.gov/html/contrib/NASA-HTTP.html (Accessed on October 1, 2016)

[19] The Slashdot Effect, hits-vs-time for the AUUG/LinuxSA InstallFest Slashdot, July 26 2000. [online] http://slash.dotat.org/~newton/installpics/slashdot-effect.html (Accessed on October 1, 2016)

[20] Deeplearning4j, open-source deep-learning library in Java, San Francisco-based business intelligence and enterprise software firm. [online] https://deeplearning4j.org (Accessed on October 1, 2016)